\documentclass[9pt, lettersize]{article}
\usepackage{amsmath,amsfonts, physics}
\usepackage{algorithmic}
\usepackage{array}
\usepackage{cite}
\usepackage[caption=false,font=normalsize,labelfont=sf,textfont=sf]{subfig}
\usepackage{textcomp}
\usepackage{stfloats}
\usepackage{url}
\usepackage{verbatim}
\usepackage{graphicx}
\usepackage[labelfont=bf]{caption}
\usepackage{balance}
\usepackage[margin=1in]{geometry}
\title{Mixed Fields Formulation for Electromagnetic Waves Confined in Dielectric Rings}
\author{Ergun Simsek, Alioune Niang, Raonaqul Islam, Logan Courtright, Gary Carter, Curtis R. Menyuk}

\begin{document}
\maketitle

\begin{abstract}
This paper presents a complete formulation for analyzing electromagnetic wave propagation in dielectric ring structures with cylindrical symmetry. We develop a full-vectorial formulation based on Maxwell's equations in cylindrical coordinates, employing three distinct approaches: E-field, H-field, and mixed E-H formulations. The method leverages the azimuthal periodicity of resonant modes, expressing field dependencies as $e^{-jm\phi}$, where $m$ is the azimuthal mode number. The governing equations are discretized on a two-dimensional $\rho$-$z$ grid using finite differences and cast as an eigenvalue problem to determine the propagation constant $\beta$ and field distributions. The proposed solver efficiently handles inhomogeneous backgrounds, including substrates and thin films, while maintaining cylindrical symmetry. Validation against commercial software (COMSOL Multiphysics) demonstrates excellent agreement across multiple test cases, including Si$_3$N$_4$ rings in SiO$_2$, rings on substrates, and torus structures, with differences in effective refractive indices typically below 0.3\%. The method provides an accurate and versatile tool for designing and optimizing integrated photonic devices based on dielectric resonators.
\end{abstract}
\section{Introduction}\label{section1}
Dielectric ring resonators are foundational components in modern photonics, serving as critical elements in a wide array of applications including optical filters, sensors, lasers, and wavelength division multiplexing systems \cite{ures_apps1, ures_apps2, switching}. Their operation relies on the principle of total internal reflection, which confines light within a high-index dielectric structure, allowing it to circulate and constructively interfere at specific resonant wavelengths. The key performance metrics of these devices—such as the quality factor (Q-factor), free spectral range, and modal volume—are directly governed by their resonant modes and the associated electromagnetic field distributions \cite{onestopshop, Die_Ring_Solver}.

Accurately modeling these resonant modes is a complex electromagnetic problem. The cylindrical symmetry of a ring resonator suggests the use of cylindrical coordinates, where the azimuthal dependence of the fields can be described by a harmonic function \( e^{-jm\phi} \), with the integer azimuthal mode number \( m \) being a critical eigenvalue of the system. While analytical solutions exist for idealized structures, practical ring resonators often feature non-homogeneous backgrounds, such as substrates or thin films, which break simple analytical symmetry and necessitate robust numerical techniques.

Several numerical methods have been employed to analyze dielectric resonators, including the Finite Element Method (FEM) \cite{FEM, Kakihara2006, SIMFEM, bane} and Finite-Difference Time-Domain (FDTD) method \cite{FDTD, Rahman2008}. Each approach has its trade-offs between accuracy, computational cost, and ease of implementation. A particular challenge lies in the formulation of the governing vector wave equations in a manner that efficiently handles material interfaces and enforces boundary conditions, especially when the permittivity \( \varepsilon \) is a function of spatial coordinates.

This report presents a comprehensive numerical formulation for analyzing the resonant modes of dielectric ring resonators embedded in cylindrically symmetric environments. We develop a full-vectorial solver based on a two-dimensional eigenproblem derived from Maxwell's equations in the frequency domain. The core of our approach involves three distinct formulations:
\begin{enumerate}
    \item An \textbf{E-field formulation}, derived directly from the vector wave equation for the electric field.
    \item An \textbf{H-field formulation}, derived from the wave equation for the magnetic field.
    \item A \textbf{mixed E-H formulation}, which couples the electric and magnetic field components to facilitate a more direct enforcement of boundary conditions at material interfaces.
\end{enumerate}

We discretize the resulting equations on a two-dimensional \( \rho-z \) grid using a finite-difference scheme, transforming the problem into a linear eigenvalue problem that is solved to find the propagation constant \( \beta \) and the corresponding field profiles. The formulation is validated against commercial software (COMSOL Multiphysics) for several benchmark cases, including a buried Si\(_3\)N\(_4\) ring, a ring on a substrate, and a torus structure, demonstrating excellent agreement and confirming the method's accuracy and versatility for designing and analyzing integrated photonic devices.
\section{Numerical Formulation}\label{section2}
\begin{figure}[h]
    \centering
    \includegraphics[width=0.8\linewidth]{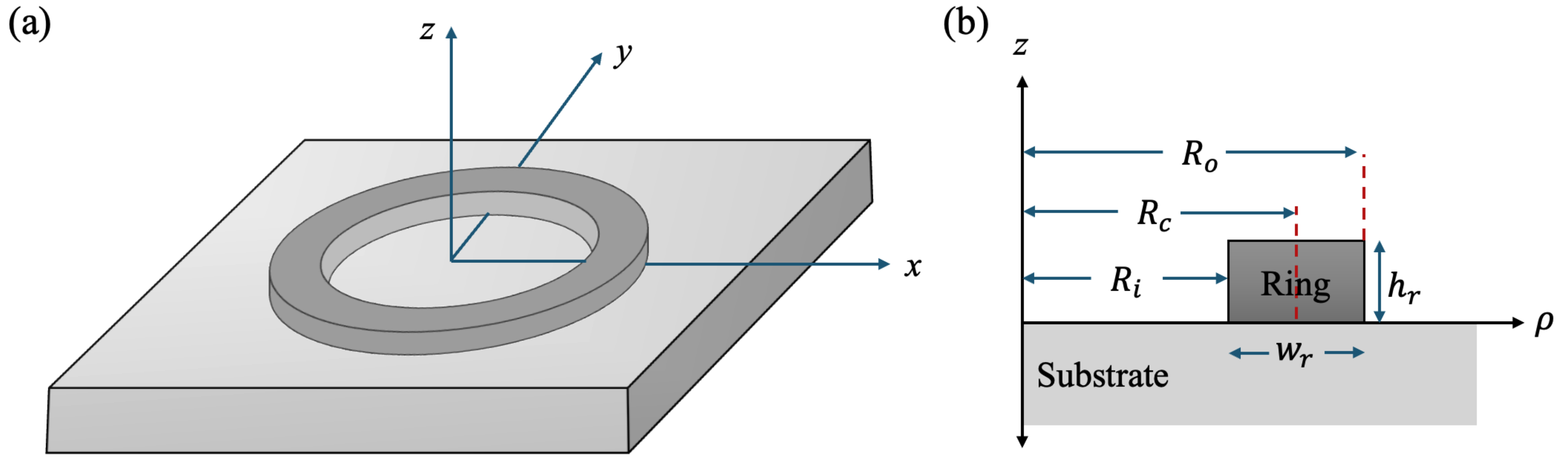}
    \caption{A dielectric ring with a width of $w_r$, a height of $h_r$, and a central radius of $R_c$ on a substrate: (a) three-dimensional and (b) two-dimensional views. $R_i$ and $R_o$ are the inner and outer radii of the ring, i.e., $R_i = R_c -w_r/2 = R_o - w_r$. By assuming the material properties are uniform axially around the $z$-axis, the problem can be solved in two dimensions using differential vector operators in the cylindrical coordinate system.}
    \label{fig:ring}
\end{figure}

Assume, we have a dielectric ring with a central radius of $R_c$, width of $w_r$, and height of $h_r$. The background does not have to be a homogeneous medium. The ring can be placed on a substrate, as illustrated in Fig. \ref{fig:ring}. The only requirement regarding the background is that it has to have cylindrical symmetry with respect to the $z$-axis. So, the relative permittivity ($\varepsilon$) is a function of $\rho$ and $z$.  We assume the entire structure is non-magnetic, i.e., $\mu(\rho,z) = \mu_0$, where $\mu_0$ is the magnetic permeability of free space. 

When they are in a resonant mode, the electromagnetic waves confined in the ring should have an $e^{-jm\phi}$ dependence, where $m = \beta R_c$ is an integer representing the azimuthal mode order and $\beta$ is the propagation constant. Due to the cylindrical symmetry, the electric and magnetic fields can be represented with 
\begin{equation}
\mathbf{E}(\rho, \phi, z) = \left\{ \hat{\rho} E_\rho(\rho, z) +\hat{\phi} E_\phi(\rho, z) + \hat{z}E_z(\rho, z) \right\} e^{-jm\phi}, 
\label{Efield}
\end{equation}
\begin{equation}
\mathbf{H}(\rho, \phi, z) = \left\{ \hat{\rho} H_\rho(\rho, z) +\hat{\phi} H_\phi(\rho, z) + \hat{z}H_z(\rho, z) \right\} e^{-jm\phi}, 
\label{Efield}
\end{equation}
where $\rho$ is the radial distance from the origin to the point projected onto the $xy$ plane, $\phi$ is the azimuthal angle, $z$ is the height or vertical distance of from the $xy$ plane.

To derive the Helmholtz wave equations, we start with Maxwell's equations for the electric and magnetic fields in a source-free, lossless, and non-magnetic medium: 
\begin{eqnarray} 
\nabla \times \mathbf{E} &= -j\omega\mu_0\mathbf{H}, \label{Max1} \\ 
\nabla \times \mathbf{H} &= j\omega\varepsilon\mathbf{E}, \label{Max2} \\ 
\nabla \cdot (\varepsilon\mathbf{E}) &= 0 \label{Max3}, \\ 
\nabla \cdot \mathbf{H} &= 0 \label{Max4}, 
\end{eqnarray} 
where $\varepsilon$ is the electrical permittivity, and $\mu_0$ is the magnetic permeability of vacuum. 

\subsection{$\mathbf{E}$-field Formulation}
Taking the curl of Eq. \eqref{Max1} and substituting Eqs. \eqref{Max2}, we obtain 
\begin{equation} \nabla \times \nabla \times \mathbf{E} = \omega^2\mu_0\varepsilon\mathbf{E}. \end{equation} 
An important point is that the divergence of electric flux is equal to zero, but $\nabla\cdot\mathbf{E} \neq 0 $. We can derive from Eq. \eqref{Max3} that 
\begin{equation} 
\nabla\cdot\mathbf{E} = - \frac{1}{\varepsilon}\nabla\varepsilon \cdot \mathbf{E} \label{divE} 
\end{equation} 
Using the vector identity $\nabla \times \nabla \times \mathbf{E} = \nabla(\nabla \cdot \mathbf{E}) - \nabla^2\mathbf{E}$ and Eq. \eqref{divE}, we obtain 
\begin{equation} 
\nabla^2\mathbf{E} + \nabla\left(\frac{1}{\varepsilon}\nabla\varepsilon \cdot \mathbf{E}\right) + k_0^2\varepsilon_r\mathbf{E} = 0, 
\label{waveeq} 
\end{equation} 
where $k_0 = \omega\sqrt{\mu_0\varepsilon_0}$ is the free-space wavenumber and $\varepsilon_r = \varepsilon/\varepsilon_0$ is the relative permittivity. \\
The $\nabla^2\mathbf{E}$ in cylindrical coordinates is given by the following equation
\begin{equation} 
\begin{aligned} 
\nabla^2\mathbf{E} = & \left[ \hat{\rho}\left( \frac{\partial^2 E_\rho}{\partial \rho^2} + \frac{1}{\rho}\frac{\partial E_\rho}{\partial \rho} + \frac{\partial^2 E_\rho}{\partial z^2} - \frac{E_\rho}{\rho^2} - \frac{m^2}{\rho^2}E_\rho  + \frac{2jm}{\rho^2}E_\phi \right) \right. \\
& 
+ \hat{\phi}\left( \frac{\partial^2 E_\phi}{\partial \rho^2} + \frac{1}{\rho}\frac{\partial E_\phi}{\partial \rho}  + \frac{\partial^2 E_\phi}{\partial z^2} - \frac{E_\phi}{\rho^2} - \frac{m^2}{\rho^2}E_\phi  - \frac{2jm}{\rho^2}E_\rho \right) \\
& \left. + \hat{z}\left(\frac{\partial^2 E_z}{\partial \rho^2} + \frac{1}{\rho}\frac{\partial E_z}{\partial \rho} + \frac{\partial^2 E_z}{\partial z^2} - \frac{m^2}{\rho^2}E_z \right) \right] e^{-jm\phi}. 
\end{aligned} 
\label{part1} 
\end{equation} 
Let's define a new operator, which will simplify our formulation
\[\mathcal{L}_0 = \frac{\partial^2}{\partial \rho^2}  + \frac{1}{\rho}\frac{\partial}{\partial \rho} + \frac{\partial^2}{\partial z^2} \]
With this new operator
\begin{equation} 
\begin{aligned} 
\nabla^2\mathbf{E} = & \left[ \hat{\rho}\left( \mathcal{L}_0\{E_\rho \} - \frac{E_\rho}{\rho^2} - \frac{m^2}{\rho^2}E_\rho  + \frac{2jm}{\rho^2}E_\phi \right) \right. \\
& 
+ \hat{\phi}\left(  \mathcal{L}_0\{E_\phi \} - \frac{E_\phi}{\rho^2} - \frac{m^2}{\rho^2}E_\phi  - \frac{2jm}{\rho^2}E_\rho \right) \\
& \left. + \hat{z}\left(  \mathcal{L}_0\{E_z \} - \frac{m^2}{\rho^2}E_z \right) \right] e^{-jm\phi}. 
\end{aligned} 
\label{part1v2} 
\end{equation} 
The $\nabla\left(\frac{1}{\varepsilon}\nabla\varepsilon \cdot \mathbf{E}\right)$ term can be expanded in three steps: we first derive an expression for $\nabla\varepsilon$, then another one for $\frac{1}{\varepsilon}\nabla\varepsilon \cdot \mathbf{E}$, and we take its gradient.
\begin{equation}
\nabla\varepsilon = \hat{\rho}\frac{\partial \varepsilon}{\partial \rho} + \hat{z}\frac{\partial \varepsilon}{\partial z}
\label{gradeps}
\end{equation}
\begin{equation}
\frac{1}{\varepsilon}\nabla\varepsilon \cdot \mathbf{E} = \frac{1}{\varepsilon} \frac{\partial \varepsilon}{\partial \rho} E_\rho + \frac{1}{\varepsilon} \frac{\partial \varepsilon}{\partial z} E_z
\end{equation}
Hence, the $\nabla\left(\frac{1}{\varepsilon}\nabla\varepsilon \cdot \mathbf{E}\right)$ term can be expanded as follows
\begin{equation}
\begin{aligned}
\nabla \left( \frac{1}{\varepsilon} \nabla \varepsilon \cdot \mathbf{E} \right) & =  \nabla \left( \frac{1}{\varepsilon} \frac{\partial \varepsilon}{\partial \rho} E_\rho e^{-jm\phi} + \frac{1}{\varepsilon} \frac{\partial \varepsilon}{\partial z} E_z e^{-jm\phi} \right) \\
 & = \hat{\rho}\left[ \frac{\partial}{\partial \rho} \left(  \frac{1}{\varepsilon} \frac{\partial \varepsilon}{\partial \rho} E_\rho + \frac{1}{\varepsilon} \frac{\partial \varepsilon}{\partial z} E_z \right) \right] e^{-jm\phi}  \\
 \ \ & + \hat{\phi}\left[ -\frac{jm}{\rho \varepsilon} \frac{\partial \varepsilon}{\partial \rho} E_\rho -\frac{jm}{\rho \varepsilon} \frac{\partial \varepsilon}{\partial z} E_z  \right]  e^{-jm\phi} \\
\  \ & + \hat{z} \left[\frac{\partial}{\partial z} \left(  \frac{1}{\varepsilon} \frac{\partial \varepsilon}{\partial \rho} E_\rho + \frac{1}{\varepsilon} \frac{\partial \varepsilon}{\partial z} E_z \right) \right] e^{-jm\phi} 
 \end{aligned}
 \end{equation}
The last term $k_0^2\varepsilon_r\mathbf{E}$ is straightforward, i.e., 
\begin{equation} 
k_0^2\varepsilon_r\mathbf{E} = k_0^2\varepsilon_r(\hat{\rho}E_\rho + \hat{z}E_z + \hat{\phi}E_\phi)e^{-jm\phi}. 
\label{part3} 
\end{equation}

Let's put everything together
\begin{equation} 
\begin{aligned} 
\nabla^2\mathbf{E} = & \left[ \hat{\rho}\left( \mathcal{L}_0\{E_\rho \} - \frac{E_\rho}{\rho^2} + \frac{m^2}{\rho^2}E_\rho  - \frac{2jm}{\rho^2}E_\phi \right) \right. \\
& 
+ \hat{\phi}\left(  \mathcal{L}_0\{E_\phi \} - \frac{E_\phi}{\rho^2} - \frac{m^2}{\rho^2}E_\phi  - \frac{2jm}{\rho^2}E_\rho \right) \\
& \left. + \hat{z}\left(  \mathcal{L}_0\{E_z \} - \frac{m^2}{\rho^2}E_z \right) \right] e^{-jm\phi}. 
\end{aligned} 
\label{part1v2} 
\end{equation} 
\begin{equation}
\nabla \left( \frac{1}{\varepsilon} \nabla \varepsilon \cdot \mathbf{E} \right) = 
\hat{\rho}\left[ \frac{\partial}{\partial \rho} \left(  \frac{1}{\varepsilon} \frac{\partial \varepsilon}{\partial \rho} E_\rho + \frac{1}{\varepsilon} \frac{\partial \varepsilon}{\partial z} E_z \right) \right]
+ \hat{\phi}\left[ -\frac{jm}{\rho \varepsilon} \frac{\partial \varepsilon}{\partial \rho} E_\rho -\frac{jm}{\rho \varepsilon} \frac{\partial \varepsilon}{\partial z} E_z  \right] 
+ \hat{z} \left[\frac{\partial}{\partial z} \left(  \frac{1}{\varepsilon} \frac{\partial \varepsilon}{\partial \rho} E_\rho + \frac{1}{\varepsilon} \frac{\partial \varepsilon}{\partial z} E_z \right)
 \right]
 \end{equation}
\begin{equation} 
k_0^2\varepsilon_r\mathbf{E} = k_0^2\varepsilon_r(\hat{\rho}E_\rho + \hat{z}E_z + \hat{\phi}E_\phi)e^{-jm\phi}. 
\label{part3} 
\end{equation}
$\hat{\rho}$-component
\[   \mathcal{L}_0\{E_\rho \} - \frac{E_\rho}{\rho^2}   + \frac{2jm}{\rho^2}E_\phi 
+ \frac{\partial}{\partial \rho} \left(  \frac{1}{\varepsilon} \frac{\partial \varepsilon}{\partial \rho} E_\rho + \frac{1}{\varepsilon} \frac{\partial \varepsilon}{\partial z} E_z \right)   +  k_0^2\varepsilon_r E_\rho = \frac{m^2}{\rho^2}E_\rho \]
$\hat{\phi}$-component
\[  \mathcal{L}_0\{E_\phi \} - \frac{E_\phi}{\rho^2}   -  \frac{2jm}{\rho^2}E_\rho 
 -\frac{jm}{\rho \varepsilon} \frac{\partial \varepsilon}{\partial \rho} E_\rho -\frac{jm}{\rho \varepsilon} \frac{\partial \varepsilon}{\partial z} E_z +  k_0^2\varepsilon_r E_\phi = \frac{m^2}{\rho^2}E_\phi \]
$\hat{z}$-component
\[ \mathcal{L}_0\{E_z \} +  \frac{\partial}{\partial z} \left(  \frac{1}{\varepsilon} \frac{\partial \varepsilon}{\partial \rho} E_\rho + \frac{1}{\varepsilon} \frac{\partial \varepsilon}{\partial z} E_z \right) +  k_0^2\varepsilon_r E_z = \frac{m^2}{\rho^2}E_z\]

\noindent Let's organize \\
$\hat{\rho}$-component
\[   \mathcal{L}_0\{E_\rho \} - \frac{E_\rho}{\rho^2} + \frac{\partial}{\partial \rho}  \frac{1}{\varepsilon} \frac{\partial \varepsilon}{\partial \rho} E_\rho +  k_0^2\varepsilon_r E_\rho + \frac{2jm}{\rho^2}E_\phi 
+ \frac{\partial}{\partial \rho}\frac{1}{\varepsilon} \frac{\partial \varepsilon}{\partial z} E_z  = \frac{m^2}{\rho^2}E_\rho \]
%
$\hat{\phi}$-component
\[ \left( -\frac{2jm}{\rho^2} -\frac{jm}{\rho \varepsilon} \frac{\partial \varepsilon}{\partial \rho} \right) E_\rho +
 \left( \mathcal{L}_0 - \frac{1}{\rho^2} +  k_0^2\varepsilon_r  \right) E_\phi -\frac{jm}{\rho \varepsilon} \frac{\partial \varepsilon}{\partial z} E_z = \frac{m^2}{\rho^2}E_\phi \]
%
$\hat{z}$-component
\[ \frac{\partial}{\partial z} \frac{1}{\varepsilon} \frac{\partial \varepsilon}{\partial \rho} E_\rho + \mathcal{L}_0\{E_z \} +  \frac{\partial}{\partial z}  \frac{1}{\varepsilon} \frac{\partial \varepsilon}{\partial z} E_z +  k_0^2\varepsilon_r E_z = \frac{m^2}{\rho^2}E_z\]

$\hat{\rho}$-component
\[  \mathcal{L}\{E_\rho \} - \frac{E_\rho}{R_c^2} + \frac{\rho^2}{R_c^2}\frac{\partial}{\partial \rho}  \frac{1}{\varepsilon} \frac{\partial \varepsilon}{\partial \rho} E_\rho  + \frac{2jm}{R_c^2}E_\phi 
+ \frac{\rho^2}{R_c^2} \frac{\partial}{\partial \rho}\frac{1}{\varepsilon} \frac{\partial \varepsilon}{\partial z} E_z  = \beta^2 E_\rho \]
%
$\hat{\phi}$-component
\[ \left( -\frac{2jm}{R_c^2} -\frac{jm \rho}{R_c^2 \varepsilon} \frac{\partial \varepsilon}{\partial \rho} \right) E_\rho +
 \left( \mathcal{L} - \frac{1}{R_c^2}  \right) E_\phi -\frac{jm\rho}{R_c^2 \varepsilon} \frac{\partial \varepsilon}{\partial z} E_z = \beta^2 E_\phi \]
%
$\hat{z}$-component
\[ \frac{\rho^2}{R_c^2}\frac{\partial}{\partial z} \frac{1}{\varepsilon} \frac{\partial \varepsilon}{\partial \rho} E_\rho + \mathcal{L}\{E_z \} +  \frac{\rho^2}{R_c^2}\frac{\partial}{\partial z}  \frac{1}{\varepsilon} \frac{\partial \varepsilon}{\partial z} E_z = \beta^2 E_z\]

Let's define a new operator, which will simplify our formulation
\[\mathcal{L} = \frac{\rho^2}{R_c^2} \left( \frac{\partial^2}{\partial \rho^2}  + \frac{1}{\rho}\frac{\partial}{\partial \rho} + \frac{\partial^2}{\partial z^2} +k_0^2 \varepsilon_r \right)  \]

From Eq. \eqref{waveeq}, we create three sets of equations by enforcing each sides of the equation to be the same in the $\hat{\rho}$, $\hat{\phi}$, and $\hat{z}$ directions. By multiplying both sides of all equations with $\rho^2/R_c^2$ and using this new operator, we obtain 
$\hat{\rho}$-component
\begin{equation}
\left( \mathcal{L} - \frac{1}{R_c^2} + \frac{\rho^2}{R_c^2}\frac{\partial}{\partial \rho}  \frac{1}{\varepsilon} \frac{\partial \varepsilon}{\partial \rho}\right) E_\rho  + \frac{2jm}{R_c^2}E_\phi 
+ \frac{\rho^2}{R_c^2} \frac{\partial}{\partial \rho}\frac{1}{\varepsilon} \frac{\partial \varepsilon}{\partial z} E_z  = \beta^2 E_\rho 
\label{EformRho}
\end{equation}
%
$\hat{\phi}$-component
\begin{equation} 
\left( -\frac{2jm}{R_c^2} -\frac{jm \rho}{R_c^2 \varepsilon} \frac{\partial \varepsilon}{\partial \rho} \right) E_\rho +
 \left( \mathcal{L} - \frac{1}{R_c^2}  \right) E_\phi -\frac{jm\rho}{R_c^2 \varepsilon} \frac{\partial \varepsilon}{\partial z} E_z = \beta^2 E_\phi 
\end{equation}
%
$\hat{z}$-component
\begin{equation}
\frac{\rho^2}{R_c^2}\frac{\partial}{\partial z} \frac{1}{\varepsilon} \frac{\partial \varepsilon}{\partial \rho} E_\rho + \mathcal{L} E_z  +  \frac{\rho^2}{R_c^2}\frac{\partial}{\partial z}  \frac{1}{\varepsilon} \frac{\partial \varepsilon}{\partial z} E_z = \beta^2 E_z
\label{EformZ}
\end{equation}
Remember
\[
\nabla \times \mathbf{E} = \left[ \left( \frac{-j m E_z}{\rho} - \frac{\partial E_\phi}{\partial z} \right) \hat{\rho} 
+ \left( \frac{\partial E_\rho}{\partial z} - \frac{\partial E_z}{\partial \rho} \right) \hat{\phi} 
+ \frac{1}{\rho} \left( \frac{\partial}{\partial \rho} \left( \rho E_\phi \right) + j m E_\rho \right) \hat{z} \right] e^{-j m \phi} = -j\omega\mu_0 \mathbf{H}
\]
Hence, we can determine the $\rho$, $\phi$, and $z$ components of the magnetic field with following set of equations
\begin{align}
H_\rho = & \frac{1}{j\omega\mu_0}  \left(  \frac{j m E_z}{\rho} + \frac{\partial E_\phi}{\partial z}  \right) \\
H_\phi = & \frac{1}{j\omega\mu_0}  \left( \frac{\partial E_z}{\partial \rho} - \frac{\partial E_\rho}{\partial z} \right) \label{hphi} \\
H_z = & -\frac{1}{j\omega\mu_0}  \left( \frac{1}{\rho} \frac{\partial}{\partial \rho} \left( \rho E_\phi \right) + \frac{j m}{\rho}  E_\rho \right) 
\end{align}

\subsection{$\mathbf{H}$-field Formulation}
The Eq. \eqref{Max2} can be also written in the following form
\[ \frac{1}{\varepsilon} \nabla \times \mathbf{H} = j\omega\mathbf{E} \]
When we take the curl of both sides
\[\nabla \times \left( \frac{1}{\varepsilon} \nabla \times \mathbf{H} \right) = j\omega \left( \nabla \times \mathbf{E} \right) \]
We substitute Eq. \eqref{Max1} and we obtain
\begin{equation}
    \nabla \times \left( \frac{1}{\varepsilon} \nabla \times \mathbf{H} \right) = \omega^2 \mu_0 \mathbf{H} 
    \label{waveHv1}
\end{equation}

Now let's work on the left hand side with the help of the product rule for the curl of a product:
\begin{equation}
\nabla \times \left( \frac{1}{\epsilon} \nabla \times \mathbf{H} \right) = \nabla \left( \frac{1}{\epsilon} \right) \times \nabla \times \mathbf{H} + \frac{1}{\epsilon} \nabla \times \nabla \times \mathbf{H}.
\label{lhs1}
\end{equation}

Let's start with the second part of the Eq. \eqref{lhs1}
\[\nabla \times \nabla \times \mathbf{H}  = \nabla \left(\nabla\cdot \mathbf{H} \right)- \nabla^2 \mathbf{H} \] 

Since $\nabla\cdot \mathbf{H} = 0$ for a non-magnetic and source-free medium, then the Eq. \eqref{waveHv1} becomes
\begin{equation}
\nabla \left( \frac{1}{\epsilon} \right) \times \nabla \times \mathbf{H} - \frac{1}{\epsilon} \nabla^2 \mathbf{H} -\omega^2 \mu_0 \mathbf{H} = 0 
    \label{waveHv2}
\end{equation}

To simplify things a bit, let's multiply the Eq. \eqref{waveHv2} with $-\varepsilon$
\begin{equation}
-\varepsilon \nabla \left( \frac{1}{\epsilon} \right) \times \nabla \times \mathbf{H} + \nabla^2 \mathbf{H} +\omega^2 \mu_0 \varepsilon \mathbf{H} = 0 
    \label{waveHv3}
\end{equation}

Notice that $\omega^2 \mu_0 \varepsilon = k_0^2\varepsilon_r$, 
\begin{equation}
-\varepsilon \nabla \left( \frac{1}{\epsilon} \right) \times \nabla \times \mathbf{H} + \nabla^2 \mathbf{H} + k_0^2\varepsilon_r \mathbf{H} = 0 
    \label{waveHv4}
\end{equation}

The first part of the first term of Eq. \eqref{waveHv4} can be written as
\[
-\varepsilon \nabla \left( \frac{1}{\epsilon} \right) = \frac{1}{\epsilon} \nabla \epsilon = \frac{1}{\epsilon} \left( \frac{\partial \epsilon}{\partial \rho} \hat{\rho} + \frac{\partial \epsilon}{\partial z} \hat{z} \right).
\]

The curl of \(\mathbf{H}\) in cylindrical coordinates is written as follows,
\[
\nabla \times \mathbf{H} = \left[ \left( \frac{-j m H_z}{\rho} - \frac{\partial H_\phi}{\partial z} \right) \hat{\rho} 
+ \left( \frac{\partial H_\rho}{\partial z} - \frac{\partial H_z}{\partial \rho} \right) \hat{\phi} 
+ \frac{1}{\rho} \left( \frac{\partial}{\partial \rho} \left( \rho H_\phi \right) + j m H_\rho \right) \hat{z} \right] e^{-j m \phi}
\]

Then, the first part of Eq. \eqref{waveHv4} can be expanded as
\begin{equation}
    \begin{aligned}
        \frac{1}{\epsilon} \left( \hat{\rho} \frac{\partial \epsilon}{\partial \rho} + \hat{z} \frac{\partial \epsilon}{\partial z} \right) \times (\nabla \times \mathbf{H}) = & 
        \left\{ \hat{\rho} \left[  \frac{1}{\epsilon} \frac{\partial \epsilon}{\partial z} \left( -\frac{\partial H_\rho}{\partial z} + \frac{\partial H_z}{\partial \rho} \right)
 \right] \right. \\
        & - \hat{\phi} \left[ \frac{1}{\epsilon} \left( \frac{\partial \epsilon}{\partial \rho} \frac{1}{\rho} \left( \frac{\partial}{\partial \rho} \left( \rho H_\phi \right) + j m H_\rho \right) - \frac{\partial \epsilon}{\partial z} \left( \frac{-j m H_z}{\rho} - \frac{\partial H_\phi}{\partial z} \right) \right)
  \right] \\
        & \left. \hat{z} \left[  \frac{1}{\epsilon} \frac{\partial \epsilon}{\partial \rho} \left( \frac{\partial H_\rho}{\partial z} - \frac{\partial H_z}{\partial \rho} \right)
 \right] \right\} e^{-j m \phi}
    \end{aligned}
    \label{part1}
\end{equation}

The second term the Eq. \eqref{waveHv4} is the Laplacian operator
\begin{equation} 
\begin{aligned} 
\nabla^2\mathbf{H} = & \left\{ \hat{\rho}\left( \frac{\partial^2 H_\rho}{\partial \rho^2} + \frac{1}{\rho}\frac{\partial H_\rho}{\partial \rho} - \frac{H_\rho}{\rho^2} - \frac{m^2}{\rho^2}H_\rho + \frac{\partial^2 H_\rho}{\partial z^2} + \frac{2jm}{\rho^2}H_\phi \right) \right. \\
& + \hat{z}\left(\frac{\partial^2 H_z}{\partial \rho^2} + \frac{1}{\rho}\frac{\partial H_z}{\partial \rho} - \frac{m^2}{\rho^2}H_z + \frac{\partial^2 H_z}{\partial z^2}\right) \\
& \left. + \hat{\phi}\left( \frac{\partial^2 H_\phi}{\partial \rho^2} + \frac{1}{\rho}\frac{\partial H_\phi}{\partial \rho} - \frac{H_\phi}{\rho^2} - \frac{m^2}{\rho^2}H_\phi + \frac{\partial^2 H_\phi}{\partial z^2} - \frac{2jm}{\rho^2}H_\rho \right) \right\} e^{-jm\phi}. 
\end{aligned} 
\label{part2} 
\end{equation} 

Based on the above derivations, we can obtain the following set of equations for the $\hat{\rho}$, $\hat{phi}$, and $\hat{z}$ components of Eq. \eqref{waveHv4}

\(\hat{\rho}\) component:
\[
\frac{1}{\epsilon} \frac{\partial \epsilon}{\partial z} \left( -\frac{\partial H_\rho}{\partial z} + \frac{\partial H_z}{\partial \rho} \right) 
+ \frac{\partial^2 H_\rho}{\partial \rho^2} + \frac{1}{\rho}\frac{\partial H_\rho}{\partial \rho} - \frac{H_\rho}{\rho^2} - \frac{m^2}{\rho^2}H_\rho + \frac{\partial^2 H_\rho}{\partial z^2} + \frac{2jm}{\rho^2}H_\phi
+ k_0^2\epsilon_r H_\rho = 0
\]

 \(\hat{\phi}\) component:
\[
\begin{aligned}
 & - \frac{1}{\epsilon} \frac{\partial \epsilon}{\partial \rho} \frac{1}{\rho} \left( \frac{\partial}{\partial \rho} \left( \rho H_\phi \right) + j m H_\rho \right) - \frac{1}{\epsilon}\frac{\partial \epsilon}{\partial z} \left( \frac{j m H_z}{\rho} + \frac{\partial H_\phi}{\partial z} \right) \\
& + \frac{\partial^2 H_\phi}{\partial \rho^2} + \frac{1}{\rho}\frac{\partial H_\phi}{\partial \rho} - \frac{H_\phi}{\rho^2} - \frac{m^2}{\rho^2}H_\phi + \frac{\partial^2 H_\phi}{\partial z^2} - \frac{2jm}{\rho^2}H_\rho
 + k_0^2\epsilon_r H_\phi = 0    
\end{aligned}
\]

 \(\hat{z}\) component:
\[  \frac{1}{\epsilon} \frac{\partial \epsilon}{\partial \rho} \left( \frac{\partial H_\rho}{\partial z} - \frac{\partial H_z}{\partial \rho} \right)
+ \frac{\partial^2 H_z}{\partial \rho^2} + \frac{1}{\rho}\frac{\partial H_z}{\partial \rho} - \frac{m^2}{\rho^2}H_z + \frac{\partial^2 H_z}{\partial z^2}
 + k_0^2\epsilon_r H_z = 0
\]

Let's define a new operator
\[
\mathcal{L} = \frac{\partial^2}{\partial \rho^2} + \frac{1}{\rho}\frac{\partial }{\partial \rho} + \frac{\partial^2 }{\partial z^2} + k_0^2\epsilon_r
\]
Then our equations become

\(\hat{\rho}\) component:
\[
\frac{1}{\epsilon} \frac{\partial \epsilon}{\partial z} \left( -\frac{\partial H_\rho}{\partial z} + \frac{\partial H_z}{\partial \rho} \right) 
+ \mathcal{L}\{H_\rho\} - \frac{H_\rho}{\rho^2} + \frac{2jm}{\rho^2}H_\phi = \frac{m^2}{\rho^2}H_\rho
\]

 \(\hat{\phi}\) component:
\[
\begin{aligned}
 &  -\frac{1}{\epsilon} \frac{\partial \epsilon}{\partial \rho} \frac{1}{\rho} \left( \frac{\partial}{\partial \rho} \left( \rho H_\phi \right) + j m H_\rho \right) - \frac{1}{\epsilon}\frac{\partial \epsilon}{\partial z} \left( \frac{j m H_z}{\rho} + \frac{\partial H_\phi}{\partial z} \right) \\
& + \mathcal{L}\{H_\phi\} - \frac{H_\phi}{\rho^2}  - \frac{2jm}{\rho^2}H_\rho = \frac{m^2}{\rho^2}H_\phi 
\end{aligned}
\]

 \(\hat{z}\) component:
\[  \frac{1}{\epsilon} \frac{\partial \epsilon}{\partial \rho} \left( \frac{\partial H_\rho}{\partial z} - \frac{\partial H_z}{\partial \rho} \right)
+ \mathcal{L}\{H_z\}
 = \frac{m^2}{\rho^2}H_z
\]
For the resonance condition, $m=\beta R_c$. Let's keep the $m$ terms on the left hand sides but replace the ones on the rights hands, and multiply both sidez with $\rho^2/R_c^2$

\(\hat{\rho}\) component:
\[
\frac{\rho^2}{R_c^2} \left\{
\frac{1}{\epsilon} \frac{\partial \epsilon}{\partial z} \left(-\frac{\partial H_\rho}{\partial z} + \frac{\partial H_z}{\partial \rho} \right) 
+ \mathcal{L}\{H_\rho\} - \frac{H_\rho}{\rho^2} + \frac{2jm}{\rho^2}H_\phi  \right\}= \beta^2 H_\rho
\]

 \(\hat{\phi}\) component:
\[\frac{\rho^2}{R_c^2} \left\{
 - \frac{1}{\epsilon} \frac{\partial \epsilon}{\partial \rho} \frac{1}{\rho} \left( \frac{\partial}{\partial \rho} \left( \rho H_\phi \right) + j m H_\rho \right) - \frac{1}{\epsilon}\frac{\partial \epsilon}{\partial z} \left( \frac{j m H_z}{\rho} + \frac{\partial H_\phi}{\partial z} \right) \\
 + \mathcal{L}\{H_\phi\} - \frac{H_\phi}{\rho^2}  - \frac{2jm}{\rho^2}H_\rho \right\} = \beta^2 H_\phi 
\]

 \(\hat{z}\) component:
\[  \frac{\rho^2}{R_c^2} \left\{\frac{1}{\epsilon} \frac{\partial \epsilon}{\partial \rho} \left( \frac{\partial H_\rho}{\partial z} - \frac{\partial H_z}{\partial \rho} \right)
+ \mathcal{L}\{H_z\}  \right\}
 = \beta^2 H_z
\]

\(\hat{\rho}\) component:
\begin{equation}
\left(\mathcal{L} -\frac{\rho^2}{R_c^2}\frac{1}{\epsilon} \frac{\partial \epsilon}{\partial z}\frac{\partial }{\partial z} -\frac{1}{R_c^2}\right) H_\rho
+ \frac{\rho^2}{R_c^2}\frac{1}{\epsilon} \frac{\partial \epsilon}{\partial z}\frac{\partial }{\partial z} H_z
+ \frac{2jm}{\rho^2}H_\phi = \beta^2 H_\rho
\label{HformRho}
\end{equation}
 \(\hat{\phi}\) component:
\begin{equation}
-jm \left( \frac{\rho^2}{R_c^2}  \frac{1}{\epsilon} \frac{\partial \epsilon}{\partial \rho} \frac{1}{\rho}  + \frac{2}{R_c^2} \right) H_\rho
+ \left( \mathcal{L} - \frac{\rho^2}{R_c^2} \frac{1}{\epsilon} \frac{\partial \epsilon}{\partial \rho} \frac{1}{\rho}\frac{\partial}{\partial \rho}\rho - \frac{\rho^2}{R_c^2} \frac{1}{\epsilon}\frac{\partial \epsilon}{\partial z}\frac{\partial }{\partial z}  - \frac{1}{R_c^2} \right) H_\phi 
- \frac{j m  \rho}{R_c^2} \frac{1}{\epsilon}\frac{\partial \epsilon}{\partial z} H_z = \beta^2 H_\phi 
\end{equation}
 \(\hat{z}\) component:
\begin{equation}  
\frac{\rho^2}{R_c^2} \frac{1}{\epsilon} \frac{\partial \epsilon}{\partial \rho} \frac{\partial H_\rho}{\partial z}
+ \left( \mathcal{L} - \frac{\rho^2}{R_c^2}  \frac{1}{\epsilon} \frac{\partial \epsilon}{\partial \rho} \frac{\partial }{\partial \rho} \right) H_z = \beta^2 H_z
\label{HformZ}
\end{equation}
Remember
\[
\nabla \times \mathbf{H} = \left[ \left( \frac{-j m H_z}{\rho} - \frac{\partial H_\phi}{\partial z} \right) \hat{\rho} 
+ \left( \frac{\partial H_\rho}{\partial z} - \frac{\partial H_z}{\partial \rho} \right) \hat{\phi} 
+ \frac{1}{\rho} \left( \frac{\partial}{\partial \rho} \left( \rho H_\phi \right) + j m H_\rho \right) \hat{z} \right] e^{-j m \phi}
\]
and
\[\nabla \times \mathbf{H} = j\omega\varepsilon\mathbf{E}\]
Then, we can determine the electric field components

\begin{align}
E_\rho = & -\frac{1}{j \omega \epsilon} \left( \frac{\partial H_\phi}{\partial z} + \frac{j m}{\rho} H_z \right) \\
E_\phi = & \frac{1}{j \omega \epsilon} \left(\frac{\partial H_\rho}{\partial z} - \frac{\partial H_z}{\partial \rho} \right) \label{ephi} \\
E_z = & \frac{1}{j \omega \epsilon} \frac{1}{\rho} \left( \frac{\partial}{\partial \rho} \rho H_\phi + j m H_\rho \right)
\end{align}
\subsection{Mixed E-H Formulation}
It is know that mixed electric field–magnetic field formulations allow for direct enforcement of boundary conditions on both electric and magnetic fields at interfaces between materials. In single-field formulations, deriving the secondary field (electric from magnetic or vice versa) often complicates boundary condition enforcement, potentially introducing errors at material interfaces. 

We can obtained the mixed formulation by expressing $H_\phi$ in Eq. \eqref{HformRho} in terms of $E_z$ and $E_\rho$ using the Eq. \eqref{hphi} and by expressing $E_\phi$ in Eq. \eqref{EformRho} in terms of $H_z$ and $H_\rho$ using the Eq. \eqref{ephi}. \\
$\hat{\rho}$-component of the E formulation becomes
\begin{equation}  
\left(\mathcal{L} - \frac{1}{R_c^2} + \frac{\rho^2}{R_c^2}\frac{\partial}{\partial \rho}  \frac{1}{\varepsilon} \frac{\partial \varepsilon}{\partial \rho} \right) E_\rho 
 + \frac{2m}{\omega \epsilon R_c^2} \left(\frac{\partial H_\rho}{\partial z} - \frac{\partial H_z}{\partial \rho} \right) + 
 \left( \frac{\rho^2}{R_c^2} \frac{\partial}{\partial \rho}\frac{1}{\varepsilon} \frac{\partial \varepsilon}{\partial z} \right) E_z   = \beta^2 E_\rho 
\label{mixedErho}
\end{equation}  
\(\hat{\rho}\) component of the H formulation becomes
\begin{equation}  
\left(\mathcal{L} -\frac{\rho^2}{R_c^2}\frac{1}{\epsilon} \frac{\partial \epsilon}{\partial z}\frac{\partial }{\partial z} -\frac{1}{R_c^2}\right) H_\rho
+ \frac{\rho^2}{R_c^2}\frac{1}{\epsilon} \frac{\partial \epsilon}{\partial z}\frac{\partial }{\partial z} H_z + \frac{2m}{\omega\mu_0 R_c^2}  \left( \frac{\partial E_z}{\partial \rho} - \frac{\partial E_\rho}{\partial z} \right) = \beta^2 H_\rho
\label{mixedHrho}
\end{equation}  
Eqs. \eqref{mixedErho}, \eqref{mixedHrho}, \eqref{EformZ}, and \eqref{HformZ} can be casted into a matrix equation such as
\begin{equation}
\begin{bmatrix}
M_1 & M_2 & M_3 & M_4\\
M_5 & M_6 & M_7 & M_8\\
M_9 & M_{10} & M_{11} & M_{12} \\
M_{13} & M_{14} & M_{15} & M_{16} \\
\end{bmatrix}
\begin{bmatrix}
E_\rho \\
E_z \\
H_\rho \\
H_z 
\end{bmatrix}
= 
\beta^2 
\begin{bmatrix}
E_\rho \\
E_z \\
H_\rho \\
H_z 
\end{bmatrix},
\end{equation}
and solved numerically.

\begin{align}
    M_1 = & \mathcal{L} - \frac{1}{R_c^2} + \frac{\rho^2}{R_c^2}\frac{\partial}{\partial \rho}  \frac{1}{\varepsilon} \frac{\partial \varepsilon}{\partial \rho} \\
    M_2 = & \frac{\rho^2}{R_c^2} \frac{\partial}{\partial \rho}\frac{1}{\varepsilon} \frac{\partial \varepsilon}{\partial z}  \\
    M_3 = & \frac{2m}{\omega \epsilon R_c^2} \frac{\partial H_\rho}{\partial z} \\
    M_4 = & -\frac{2m}{\omega \epsilon R_c^2}  \frac{\partial H_z}{\partial \rho} \\
    M_5 = & \frac{\rho^2}{R_c^2}\frac{\partial}{\partial z} \frac{1}{\varepsilon} \frac{\partial \varepsilon}{\partial \rho}   \\
    M_6 = & \mathcal{L} +  \frac{\rho^2}{R_c^2}\frac{\partial}{\partial z}  \frac{1}{\varepsilon} \frac{\partial \varepsilon}{\partial z} \\
    M_7 = &  0 \\
    M_8 = & 0  \\
    M_9 = &  - \frac{2m}{\omega\mu_0 R_c^2} \frac{\partial}{\partial z}  \\
    M_{10} = &  \frac{2m}{\omega\mu_0 R_c^2} \frac{\partial}{\partial \rho}  \\
    M_{11} = & \mathcal{L} -\frac{\rho^2}{R_c^2}\frac{1}{\epsilon} \frac{\partial \epsilon}{\partial z}\frac{\partial }{\partial z} -\frac{1}{R_c^2}\\
    M_{12} = & \frac{\rho^2}{R_c^2}\frac{1}{\epsilon} \frac{\partial \epsilon}{\partial z}\frac{\partial }{\partial z} \\
    M_{13} = & 0  \\
    M_{14} = & 0 \\
    M_{15} = & \frac{\rho^2}{R_c^2} \frac{1}{\epsilon} \frac{\partial \epsilon}{\partial \rho} \frac{\partial }{\partial z} \\
    M_{16} = &  \mathcal{L} - \frac{\rho^2}{R_c^2}  \frac{1}{\epsilon} \frac{\partial \epsilon}{\partial \rho} \frac{\partial }{\partial \rho}
\end{align}

Note that in the above, we keep the unknown azimuthal mode number on the left-hand side, while its explicit version $m=\beta R_c$ is used for the $m^2$ terms, which are later moved to the right-hand sides.
The eigen solution of these equations will give us the right $\beta$ value, but at the moment, we do not know the value of $m$.

We propose the following two-step algorithm to determine the correct $m$ value and solve for the fields.
For the initial calculations, let's use an approximate $m$ value given by the following expression
\begin{equation}
    m_{\rm approx} = \lfloor  \xi k_0 R_c n_{\rm r} \rceil ,
    \label{eqma1}
\end{equation}
where $\xi$ is an arbitrary positive number closer to 1.0, $n_{\rm r}$ is the refractive index of the ring, and $\lfloor \ \rceil$ represents the rounding operation. The idea behind using a $\xi$ value, which is slightly smaller than 1.0, is that we know the effective index of the dielectric waveguides is slightly lower than the refractive index of the ring for the fundamental modes. We use this approximate azimuthal mode number, build our Hamiltonian matrix, and solve the eigenvalue problem. At this point, we are only interested in the $i^{th}$ eigenvalue $\beta_{\rm approx, i}$, not the eigenvectors, where $i$ is the number of the desired resonant mode. \footnote{In theory, there should be $N_\rho\times N_z$ eigenvalues, where $N_\rho$ and $N_z$ are the numbers of nodes along the $\rho$ and $z$ directions, but both because of the non-linearity of the problem under the investigation and numerical round-off errors, only a handful of these solutions would be realistic and reliable. A very simple approach to prevent spurious solutions is to eliminate all the solutions that lead to negative attenuation constants and the solutions that lead to effective refractive indices seriously larger than the refractive index of the ring.} Because in the second step, we use this $\beta_{{\rm approx},i}$ to calculate the $m_i$ value with the following expression
\begin{equation}
    m_i = \lfloor \beta_{\rm approx} R_c \rceil,
    \label{eqma2}
\end{equation}
and solve the eigenvalue problem one more time. The eigenvalue of the $i^{th}$ mode will lead to a slightly different propagation constant, $\beta_{{\rm final},i}$. When we calculate the $\beta R_c$ multiplication with $\beta_{{\rm final},i}$, the result turns out to be a positive number, which is very close to an integer. It only differs from an integer in the third or fourth digit, e.g., 267.999 or 267.9999 instead of 268. The eigenvectors of the solution are the resonant electric fields $E_\rho$, $E_z$, $H_\rho$, and $H_z$. The $\phi$ components are computed using Eqs. \eqref{ephi} and \eqref{hphi}. When all the field components are determined, we can compute the power density of the electromagnetic wave and normalize all the fields so that the power density is equal to 1 W/m$^2$.

\section{Direct Approach}
By using Eq. \eqref{ephi}, we rewrite Eq. \eqref{EformRho}
\begin{equation}  
\begin{aligned}
 & \left(\mathcal{L} - \frac{1}{R_c^2} + \frac{\rho^2}{R_c^2}\frac{\partial}{\partial \rho}  \frac{1}{\varepsilon} \frac{\partial \varepsilon}{\partial \rho} \right) E_\rho 
+ \left( \frac{\rho^2}{R_c^2} \frac{\partial}{\partial \rho}\frac{1}{\varepsilon} \frac{\partial \varepsilon}{\partial z} \right) E_z \\
 & + \frac{2\beta}{\omega \varepsilon R_c} \left(\frac{\partial H_\rho}{\partial z} - \frac{\partial H_z}{\partial \rho} \right)   = \beta^2 E_\rho 
\end{aligned}
\label{mixedErho}
\end{equation}  
Similarly,  $\hat{\rho}$ component of the $H$-formulation becomes
\begin{equation}  
\begin{aligned}
 & \left(\mathcal{L} -\frac{\rho^2}{R_c^2}\frac{1}{\varepsilon} \frac{\partial \varepsilon}{\partial z}\frac{\partial }{\partial z} -\frac{1}{R_c^2}\right) H_\rho
+ \frac{\rho^2}{R_c^2}\frac{1}{\varepsilon} \frac{\partial \varepsilon}{\partial z}\frac{\partial }{\partial z} H_z \\
 & + \frac{2\beta}{\omega\mu_0 R_c}  \left( \frac{\partial E_z}{\partial \rho} - \frac{\partial E_\rho}{\partial z} \right) = \beta^2 H_\rho
\end{aligned}
\label{mixedHrho}
\end{equation}  
Eqs. \eqref{mixedErho}, \eqref{mixedHrho}, \eqref{EformZ}, and \eqref{HformZ} can be cast into a matrix equation such as
\begin{equation}
\overline{\overline{{M}}}\mathbf{E} + \beta \overline{\overline{{L}}} \mathbf{E} - \beta^2 \overline{\overline{{I}}} \mathbf{E} = 0,
\end{equation}
where $\overline{\overline{{M}}}$ is the matrix independent of $\beta$, $\overline{\overline{{L}}}$ is the linear term in $\beta$, and the last term is the quadratic term in $\beta$, similar to \cite{Die_WG}. 

\begin{equation}
\overline{\overline{{M}}}  = 
\begin{bmatrix}
M_1 & M_2 & 0 & 0 \\
M_3 & M_4 & 0 & 0  \\
0 & 0  & M_{5} & M_{6} \\
0 & 0  & M_{7} & M_{7} \\
\end{bmatrix}
\end{equation}
\begin{align}
    M_1 = & \mathcal{L} - \frac{1}{R_c^2} + \frac{\rho^2}{R_c^2}\frac{\partial}{\partial \rho}  \frac{1}{\varepsilon} \frac{\partial \varepsilon}{\partial \rho} \\
    M_2 = & \frac{\rho^2}{R_c^2} \frac{\partial}{\partial \rho}\frac{1}{\varepsilon} \frac{\partial \varepsilon}{\partial z}  \\
    M_3 = & \frac{\rho^2}{R_c^2}\frac{\partial}{\partial z} \frac{1}{\varepsilon} \frac{\partial \varepsilon}{\partial \rho}   \\
    M_4 = & \mathcal{L} +  \frac{\rho^2}{R_c^2}\frac{\partial}{\partial z}  \frac{1}{\varepsilon} \frac{\partial \varepsilon}{\partial z} 
\end{align}
\begin{align}
    M_{5} = & \mathcal{L} -\frac{\rho^2}{R_c^2}\frac{1}{\varepsilon} \frac{\partial \varepsilon}{\partial z}\frac{\partial }{\partial z} -\frac{1}{R_c^2}\\
    M_{6} = & \frac{\rho^2}{R_c^2}\frac{1}{\varepsilon} \frac{\partial \varepsilon}{\partial z}\frac{\partial }{\partial z} \\
    M_{7} = & \frac{\rho^2}{R_c^2} \frac{1}{\varepsilon} \frac{\partial \varepsilon}{\partial \rho} \frac{\partial }{\partial z} \\
    M_{8} = &  \mathcal{L} - \frac{\rho^2}{R_c^2}  \frac{1}{\varepsilon} \frac{\partial \varepsilon}{\partial \rho} \frac{\partial }{\partial \rho}
\end{align}

\begin{equation}
\overline{\overline{{L}}}  = 
\begin{bmatrix}
0 & 0 & L_1 & L_2\\
0 &  0 & 0 & 0 \\
L_3 & L_{4} & 0 & 0 \\
0 &  0 & 0 & 0 \\
\end{bmatrix}
\end{equation}
\begin{align}
    L_1 = & \frac{2}{\omega \varepsilon R_c} \frac{\partial }{\partial z} \\
    L_2 = & -\frac{2}{\omega \varepsilon R_c}  \frac{\partial }{\partial \rho} \\
    L_3 = &  - \frac{2}{\omega\mu_0 R_c} \frac{\partial}{\partial z}  \\
    L_4 = &  \frac{2}{\omega\mu_0 R_c} \frac{\partial}{\partial \rho} 
\end{align}

 This can be rewritten as a linear generalized eigenvalue problem by introducing an auxiliary variable $\mathbf{F} = \beta \mathbf{E}$, leading to:
\begin{equation}
    \begin{cases}
        \overline{\overline{{M}}} \mathbf{E} + \overline{\overline{{L}}} \mathbf{F} - \mathbf{F} \beta = 0, \\
        \mathbf{F} = \beta \mathbf{E},
    \end{cases}
\end{equation}
 which gives the augmented system:
\begin{equation}
    \begin{pmatrix}
        \overline{\overline{{M}}} & \overline{\overline{{L}}} \\
        0 & \overline{\overline{{I}}}
    \end{pmatrix}
    \begin{pmatrix}
        \mathbf{E} \\
        \mathbf{F}
    \end{pmatrix}
    = \beta
    \begin{pmatrix}
        0 & \overline{\overline{{I}}} \\
        \overline{\overline{{I}}} & 0
    \end{pmatrix}
    \begin{pmatrix}
        \mathbf{E} \\
        \mathbf{F}
    \end{pmatrix}.
\end{equation}
We solve the linear generalized eigenvalue problem using MATLAB's \textsf{eigs} function, which is optimized for large sparse matrices. After solving, we retain only the physically meaningful modes (those with $n_{\rm eff}^{\rm ring}<n_{\rm{ring}}$, where $n_{\rm eff}^{\rm ring} = \beta/k_0$ is the effective index and $n_{\rm{ring}}$ is the refractive index of the ring.)

\section{Numerical Results}\label{section3}
We discretize the computational domain along a rectangular grid by selecting $N_\rho$ uniformly distributed samples along the $\rho$-direction and $N_z$ uniformly distributed samples along the $z$-direction. We compute the first- and second-order derivatives using central finite differences. We impose Neumann boundary conditions \cite{Leveque} by setting the normal derivatives of the field to zero at the outer boundaries of the computational domain. As explained on page 36 of \cite{kogelnik1975theory}, the fields outside decay exponentially for the modes guided in dielectric waveguides. Hence, placing the ring at the center of a computational domain that is $(w_r +\lambda) \times (h_r +\lambda)$, where $w_r$ and $h_r$ are the ring width and height, is sufficient to obtain accurate results for determining the resonant modes of a ring surrounded by a homogeneous background. However, if there is a thin film and/or a substrate, the computational domain should be enlarged to adequately capture their effect on wave propagation. Neumann boundary conditions, which enforce zero field derivatives at the domain boundaries, can affect the accuracy in such cases by potentially introducing artificial reflections if the domain is not sufficiently large. This limitation becomes more pronounced when the field's evanescent tails interact with the boundary, necessitating careful domain size selection to minimize inaccuracies. For all the examples, the excitation wavelength is 1550 nm.

\subsection{A Si$_3$N$_4$ Ring Surrounded with SiO$_2$} 
To validate the accuracy of the formulation, we start with a dielectric ring studied in \cite{Greg2019}. The ring's central radius, width, and height are 23 $\mu$m, 1.5 $\mu$m, and 0.7 $\mu$m, respectively. The refractive index of the ring and background are 1.9761 and 1.444. For the numerical solution, we use a $150\times70$ grid on the $\rho-z$ plane, where the mesh sampling density along both directions is 50 nm. 

\begin{figure}[h]
    \centering
    \includegraphics[width=0.8\linewidth]{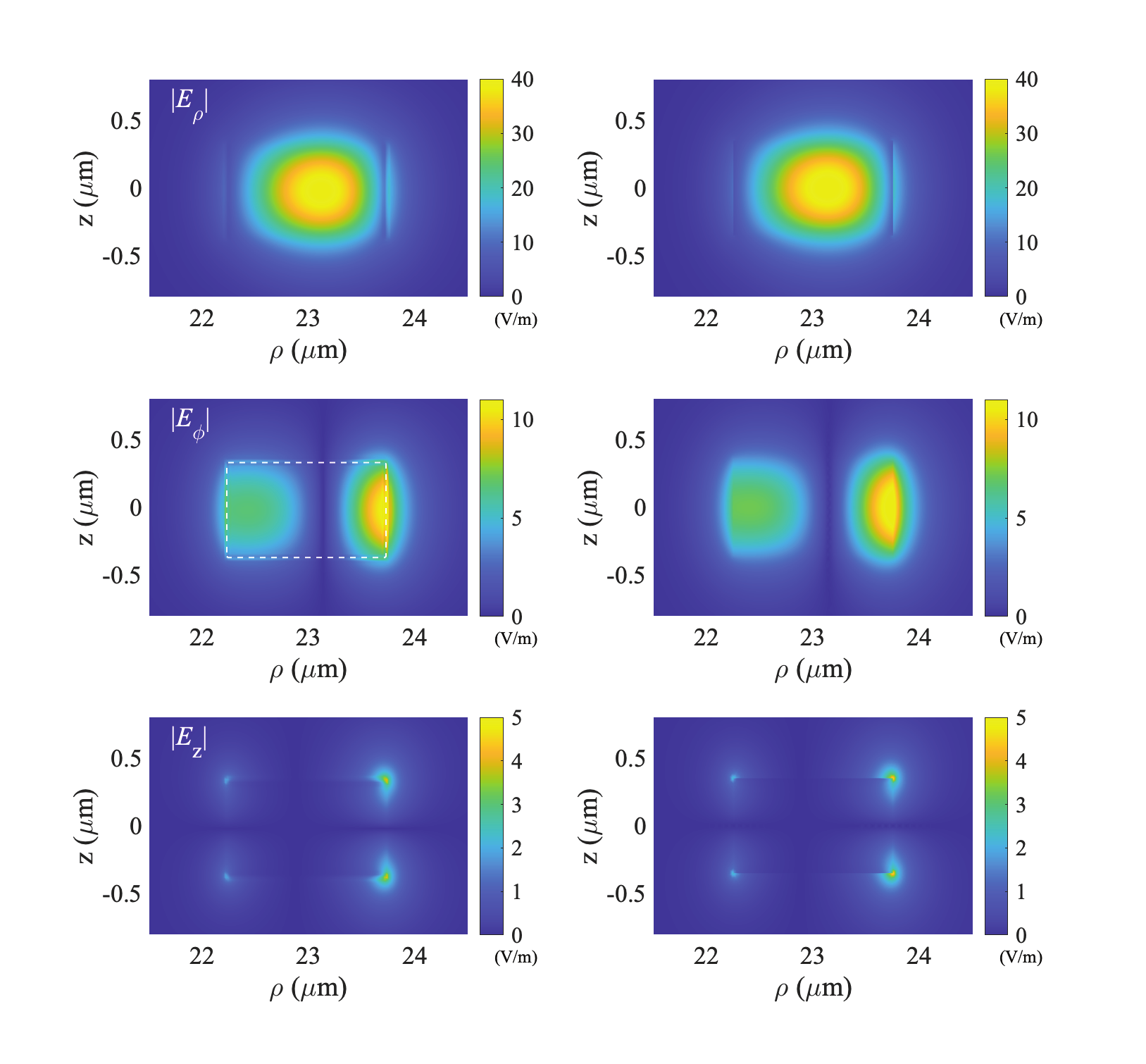}
    \caption{Magnitude of the electric field's $\rho$ (top), $\phi$ (middle), and $z$ (bottom) components for the first resonant mode of the electromagnetic waves computed with our solver (left) and COMSOL (right) in a Si$_3$N$_4$ ring with a central radius, width, and height of 23 $\mu$m, 1.5 $\mu$m, and 0.7 $\mu$m.The white dashed lines outline the boundaries of the ring, giving insight into how the electromagnetic fields are confined but not completely symmetric.}
    \label{fig:GregFields}
\end{figure}

\begin{figure}[h]
    \centering
    \includegraphics[width=\linewidth]{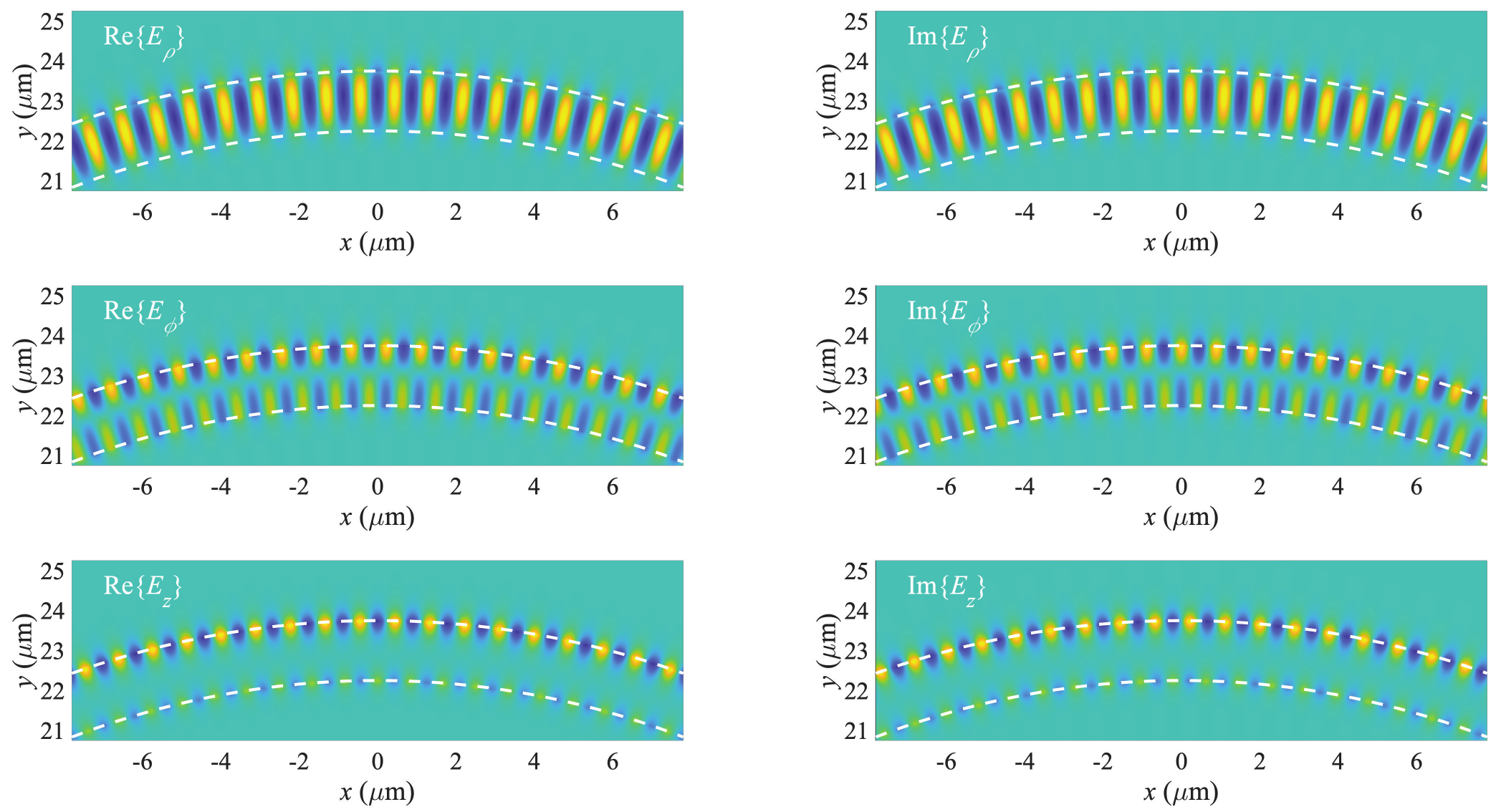}
    \caption{(Left) Real and (right) imaginary parts of the electric field's $\rho$ (top row), $\phi$ (middle row), and $z$ (bottom row) components on the $xy$-plane at $z=h_r/2$. The white dashed lines outline the inner and outer boundaries of the ring.}
    \label{fig:GregFieldsAzimuthal}
\end{figure}

Figure \ref{fig:GregFields} shows the magnitude of the electric field's $\rho$, $\phi$, and $z$ components on the $\rho-z$ plane for the first resonant mode obtained with our solver (left column) and COMSOL (right column). These solutions demonstrate a close agreement, with only minor deviations within acceptable numerical tolerance levels. We observe that the electric field is well-confined within the high-index Si$_3$N$_4$ region, with some fields extending slightly outside the boundaries due to the evanescent field. The symmetry of the fields is consistent with the fundamental resonant mode, where fields are strongly coupled to the geometry of the ring. Figure \ref{fig:GregFieldsAzimuthal} shows the real and imaginary parts of the electric field components on the $x-y$ plane at the center of the ring, i.e., at $z = h_r/2$.

Table \ref{table_greg} lists the effective refractive indices computer with COMSOL and our solver. Since we use a very fine mesh (40 points per wavelength sampling density) in the COMSOL implementation, we take COMSOL solutions as the ground truth and calculate the difference accordingly. As listed in the last column of Table \ref{table_greg}, the difference is less than 0.3 \% for all modes. 

\begin{table} 
\caption{Effective refractive indices of the first four resonant modes of a Si$_3$N$_4$ ring buried in SiO$_2$, computed with COMSOL and our solver. The fourth column is the percentage difference between COMSOL and our solutions. }\vspace{-5mm} 
\begin{center} 
\begin{tabular}{|c|c|c|c|} 
\hline 
mode index  & $n_{\mathrm{eff}}$ (COMSOL) & $n_{\mathrm{eff}}$ (This work) & Difference (\%) \\ 
\hline 
1 & 1.7909  &  1.7901  &  0.043 \\
\hline
2 & 1.7524  &  1.7491  &  0.191 \\
\hline
3 & 1.6257  &  1.6229  &  0.175 \\
\hline
4 & 1.6092  &  1.6049  &  0.265 \\
\hline
\end{tabular} 
\end{center} 
\label{table_greg} 
\end{table} 

\begin{figure}[!h]
    \centering
    \includegraphics[width=0.5\linewidth]{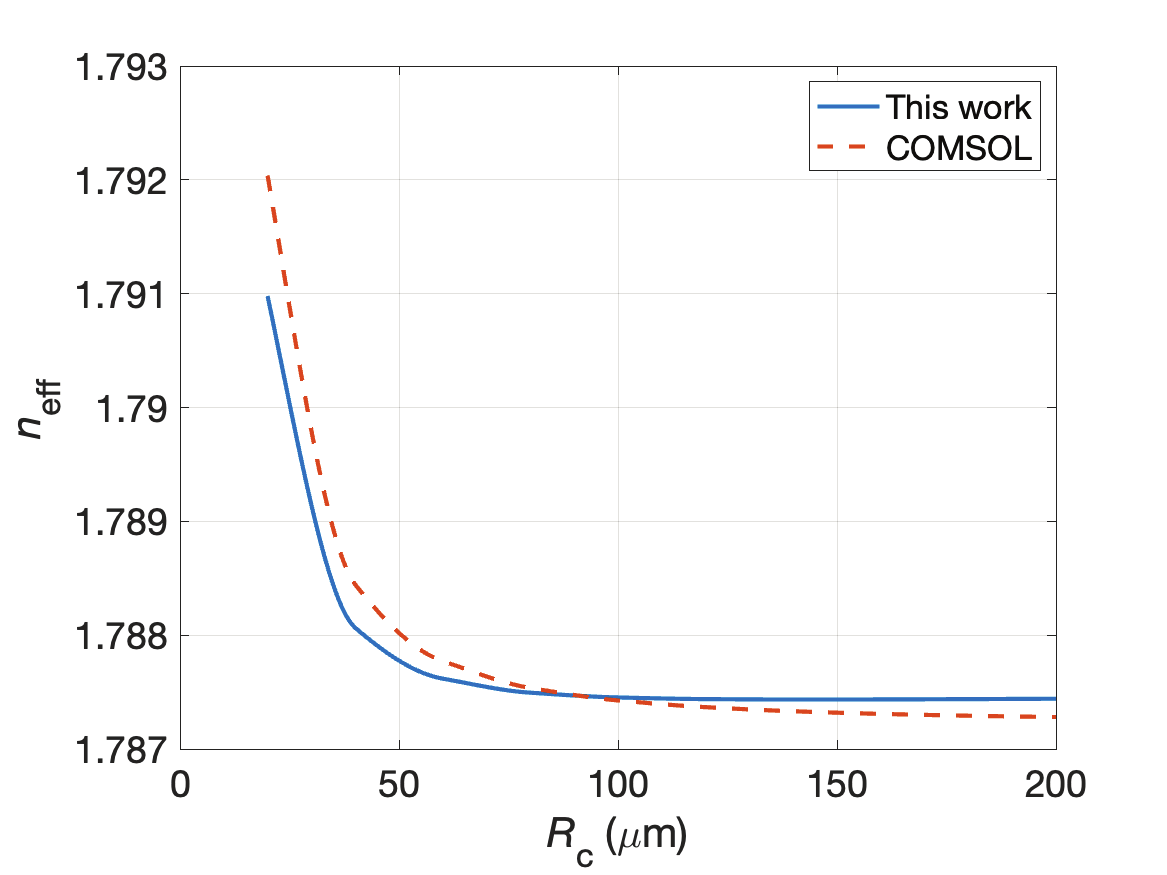}
    \caption{Effective index vs. central radius for the first resonant mode.}
    \label{fig:neff_vs_Rc}
\end{figure}

Next, we study how the ring central radius ($R_c$) affects the ring effective index. We keep all the parameters the same except the $R_c$ and calculate the effective index of the first mode as a function of $R_c$ for  $20 \ \mu\rm{m} \leq R_c \leq 200 \mu\rm{m}$. Figure \ref{fig:neff_vs_Rc} shows our and COMSOL results, where the difference is less than 0.06 \%.
The curvature significantly influences the wave confinement in smaller-radius rings, leading to a higher effective index. This is because the mode is more tightly confined, with greater overlap with high-index material. As the radius increases, the curvature effects diminish, and the ring resembles a straight waveguide. Consequently, the effective index approaches a constant value corresponding to the straight waveguide mode.

\subsection{A Si$_3$N$_4$ Ring on a Si$_3$N$_4$ Thin Film Coated Glass Substrate} 
\begin{figure}[!h]
    \centering
    \includegraphics[width=0.8\linewidth]{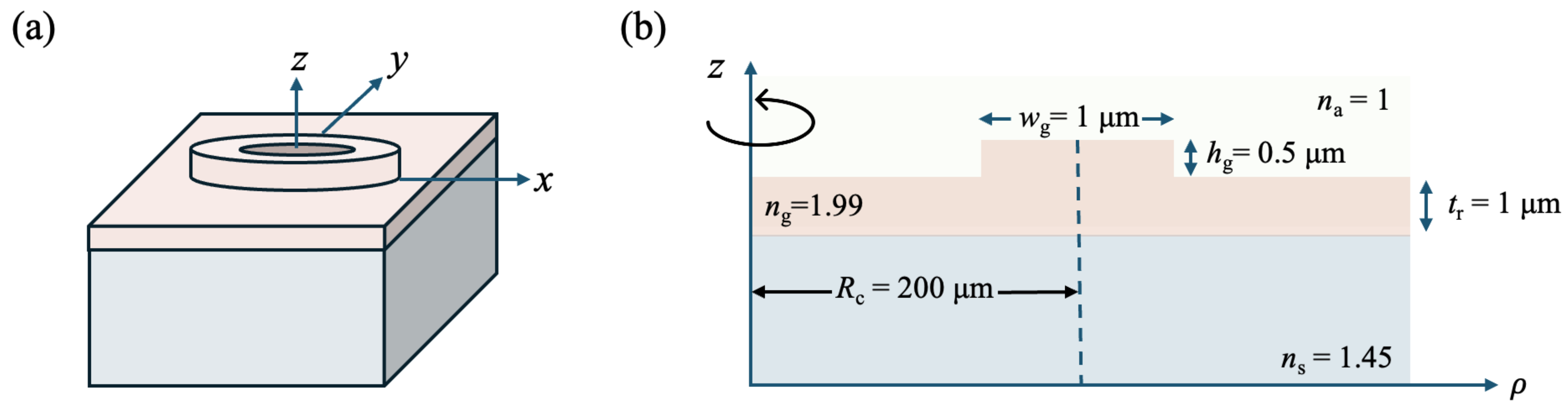}
    \caption{(a) 3D and (b) 2D views of the geometry studied in the second example: a Si$_3$N$_4$ ring grown on a Si$_3$N$_4$ thin film-coated glass substrate.}
    \label{fig:geo1}
\end{figure}

As for the second example, we work on a structure inspired by the bent waveguide studied in \cite{Prkna2004}. As shown in Fig. \ref{fig:geo1}, the structure includes a ring on a thin film-coated glass substrate. The material of the ring and thin film are the same (Si$_3$N$_4$). In \cite{Prkna2004}, they assume the refractive index of Si$_3$N$_4$ is $n_g=1.99$ at the excitation wavelength of 1.55 $\mu$m. Even though this value of $n_g$ is slightly different than the one reported in \cite{Si3N4Sellmeier}, we use it as is to be able to compare our results with the ones reported in \cite{Prkna2004}. The central radius ($R_c$), width ($w_g$), and height ($h_g$) of the ring are $200 \ \mu$m, 1 $\mu$m, and 0.5 $\mu$m, respectively. The thickness of the film ($t_r$) is 1 $\mu$m. The refractive indices of the substrate ($n_s$) and the region above the structure ($n_a$) are 1.45 and 1, respectively. For the numerical solution, we use a $120\times80$ grid on the $\rho-z$ plane, where the mesh sampling along both directions is 50 nm.

\begin{table}[!h]
\caption{Effective refractive indices of the first four resonant modes of the ring shown in Fig. \ref{fig:geo1}, computed with COMSOL and our solver. The fourth column is the percentage difference between COMSOL and our solutions. }\vspace{-5mm} 
\begin{center} 
\begin{tabular}{|c|c|c|c|} 
\hline 
mode index  & $n_{\mathrm{eff}}$ (COMSOL) & $n_{\mathrm{eff}}$ (This work) & Difference (\%) \\ 
\hline 
1 &     1.9124   &   1.9148   &  -0.127 \\
\hline
2 &     1.9041   &   1.9017   &   0.128 \\
\hline
3 &     1.8936   &   1.8913   &   0.122 \\
\hline
4 &     1.8791   &   1.8759   &   0.171 \\
\hline
\end{tabular} 
\end{center} 
\label{table_case1} 
\end{table} 

Table \ref{table_case1} compares the effective refractive index calculated with a commercial electromagnetic solver (COMSOL) and our solver for the first four resonant modes. We observe that the difference is less than 0.2 \% for all modes. Both the COMSOL solution and our solution for the fundamental mode's effective refractive are close to the value (1.911) reported in \cite{Prkna2004}. However, one should note that in \cite{Prkna2004}, the researchers deal with a bent waveguide, not a ring, and bent waveguides exhibit more substantial losses than dielectric rings.

\begin{figure}[!h]
    \centering
    \includegraphics[width=0.8\linewidth]{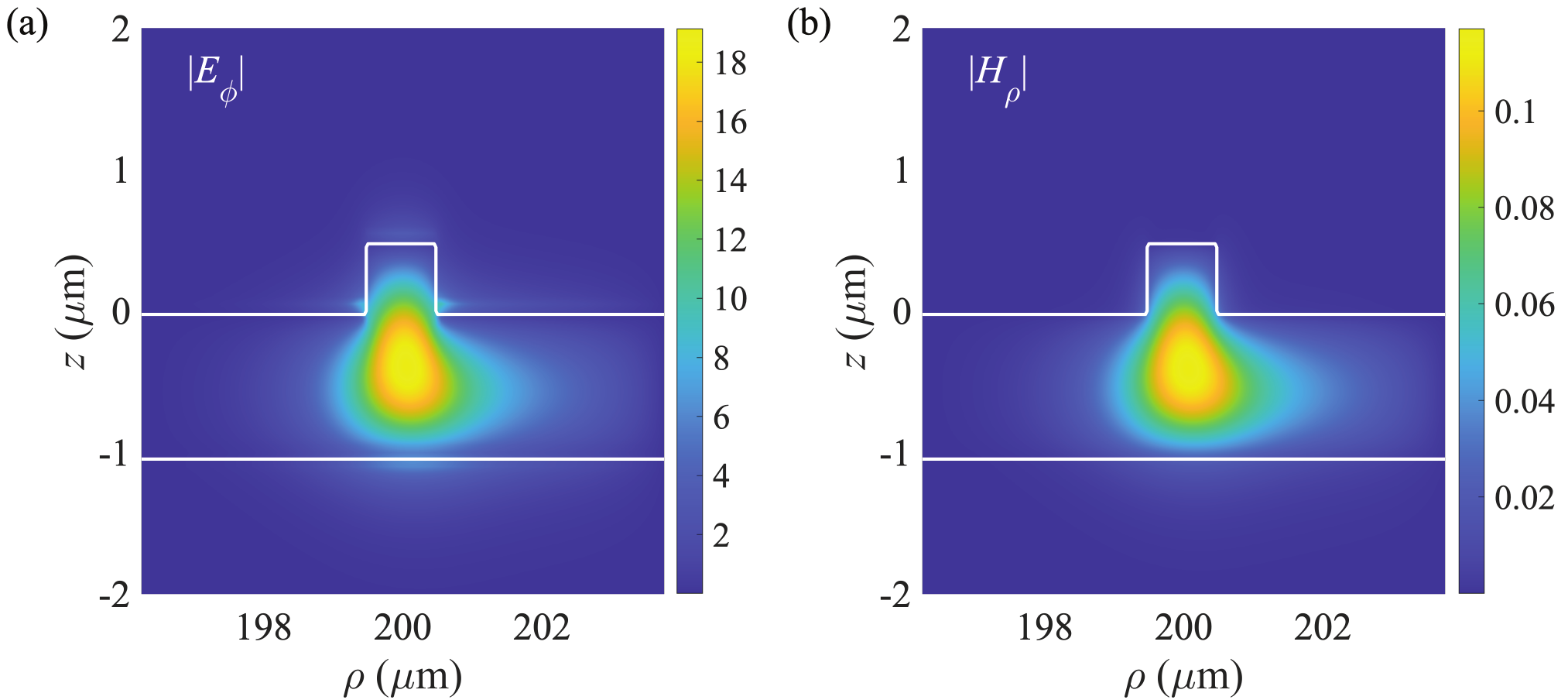}
    \caption{The magnitude of the $\phi$ component of the electric field and $\rho$ component of the magnetic field of the third resonant mode.}
    \label{fig:prkna_fields}
\end{figure}

In Fig. \ref{fig:prkna_fields}, we plot the magnitude of the $\phi$ component of the electric field and $\rho$ component of the magnetic field of the third resonant mode. Rather than complete confinement in the ring part of the structure, we observe that both fields are skewed and spread along the Si$_3$N$_4$ region but with minimal field leakage into the air and substrate regions.

\subsection{Resonant Modes of a Torus} 
\begin{figure}[!h]
    \centering
    \includegraphics[width=0.5\linewidth]{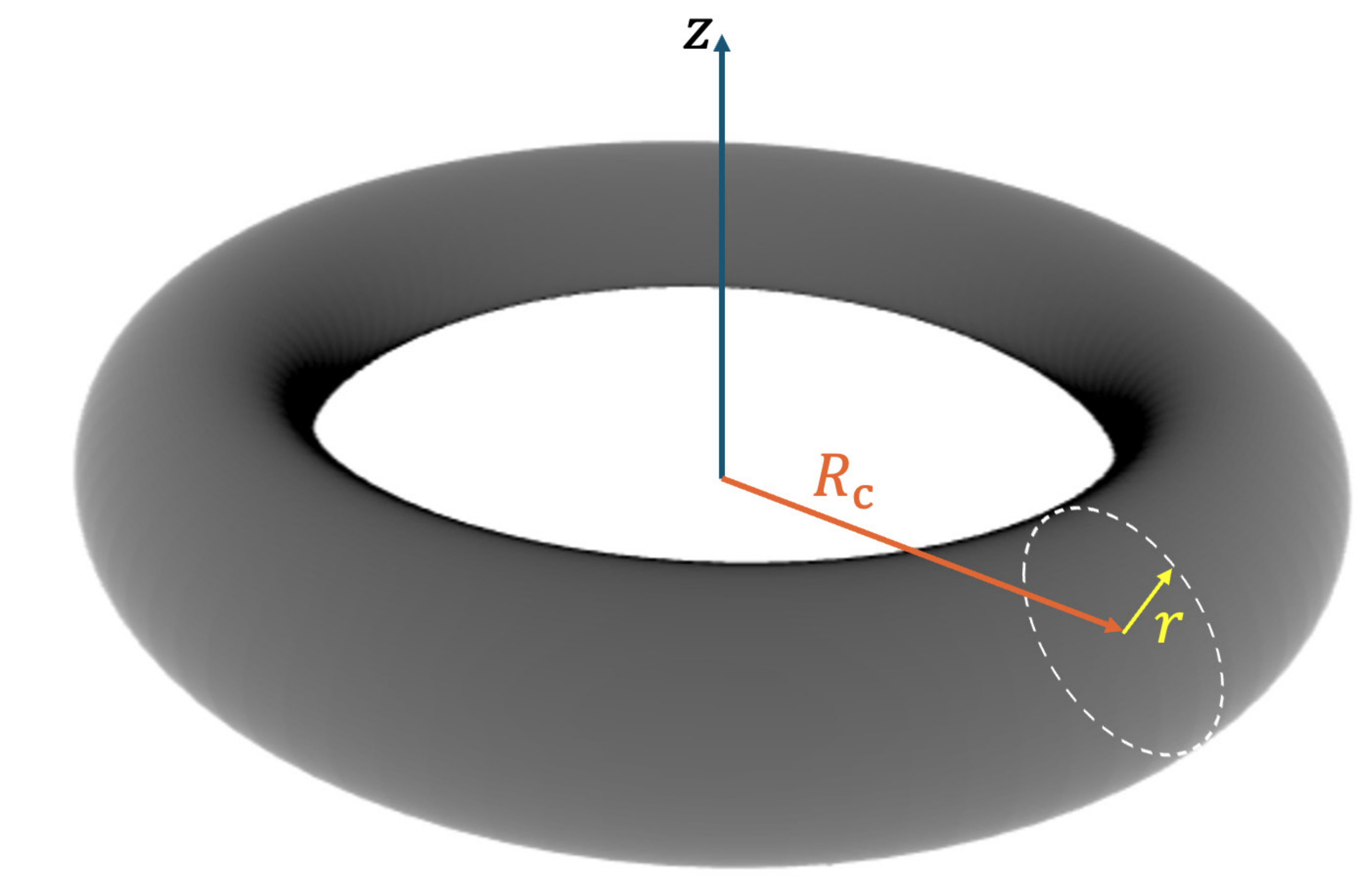}
    \caption{Illustration of a torus with a major radius of $R_c$ and a minor radius of $r$.}
    \label{fig:geo4}
\end{figure}

As for the third example, we work on the resonant modes of a torus to showcase the flexibility of the formulation. Figure \ref{fig:geo4} illustrates a torus with its major and minor radii, $R_c$ and $r$, respectively. The torus is made from Si$_3$N$_4$, and the surrounding medium is assumed to be SiO$_2$. At the wavelength of 1.55 $\mu$m, their refractive indices are calculated to be 1.9963 and 1.444 based on the formulas given in \cite{Si3N4Sellmeier} and \cite{Palik}. $R_c = 20 \ \mu$m and $r = 0.6 \ \mu$m. For the numerical solution, we use a $90\times90$ grid on the $\rho-z$ plane, using 30 nm mesh sampling density along both directions.

In Table \ref{TableTorus}, we list the effective indices of the first 11 modes determined with our solver and Tidy3D, along with the six resonant modes that we could determine with COMSOL Multiphysics and Ansoft Lumerical. 
\begin{table}[!h]
\caption{Effective indices of resonant modes in a torus determined with our implementation, Tidy3D, COMSOL Multiphysics, and Ansys Lumerical.}
\begin{center}
\begin{tabular}{|c|c|c|c|c|}
\hline
mode  & $n_{\rm{eff}}$ & $n_{\rm{eff}}$ & $n_{\rm{eff}}$ & $n_{\rm{eff}}$ \\
index & (This work) & (Tidy3D) & (COMSOL) & (Lumerical)  \\
\hline
1 & 1.8296 & 1.847553 & 1.831 & 1.83092 \\
\hline
2 & 1.8284 & 1.847363 & 1.830 & 1.83033 \\
\hline
3 & 1.6196 & 1.635399 & 1.621 & 1.6204 \\
\hline
4 & 1.5735 & 1.591508 & 1.576 & 1.5774 \\
\hline
5 & 1.561 & 1.578471 & 1.564 & 1.5645 \\
\hline
6  & 1.5572 & 1.577542 & 1.563 & 1.5623 \\
\hline
7 & 1.4577 & 1.476436 &  &  \\
\hline
8  & 1.4564 & 1.468303 &  &  \\
\hline
9 & 1.4105 & 1.416233 &  & \\
\hline
10 & 1.4077 & 1.391547 &  & \\
\hline
11 & 1.3666 & 1.3683 &  & \\
\hline
\end{tabular}
\end{center}
\label{TableTorus}
\end{table}%

\begin{figure}[h]
    \centering
    \includegraphics[width=0.8\linewidth]{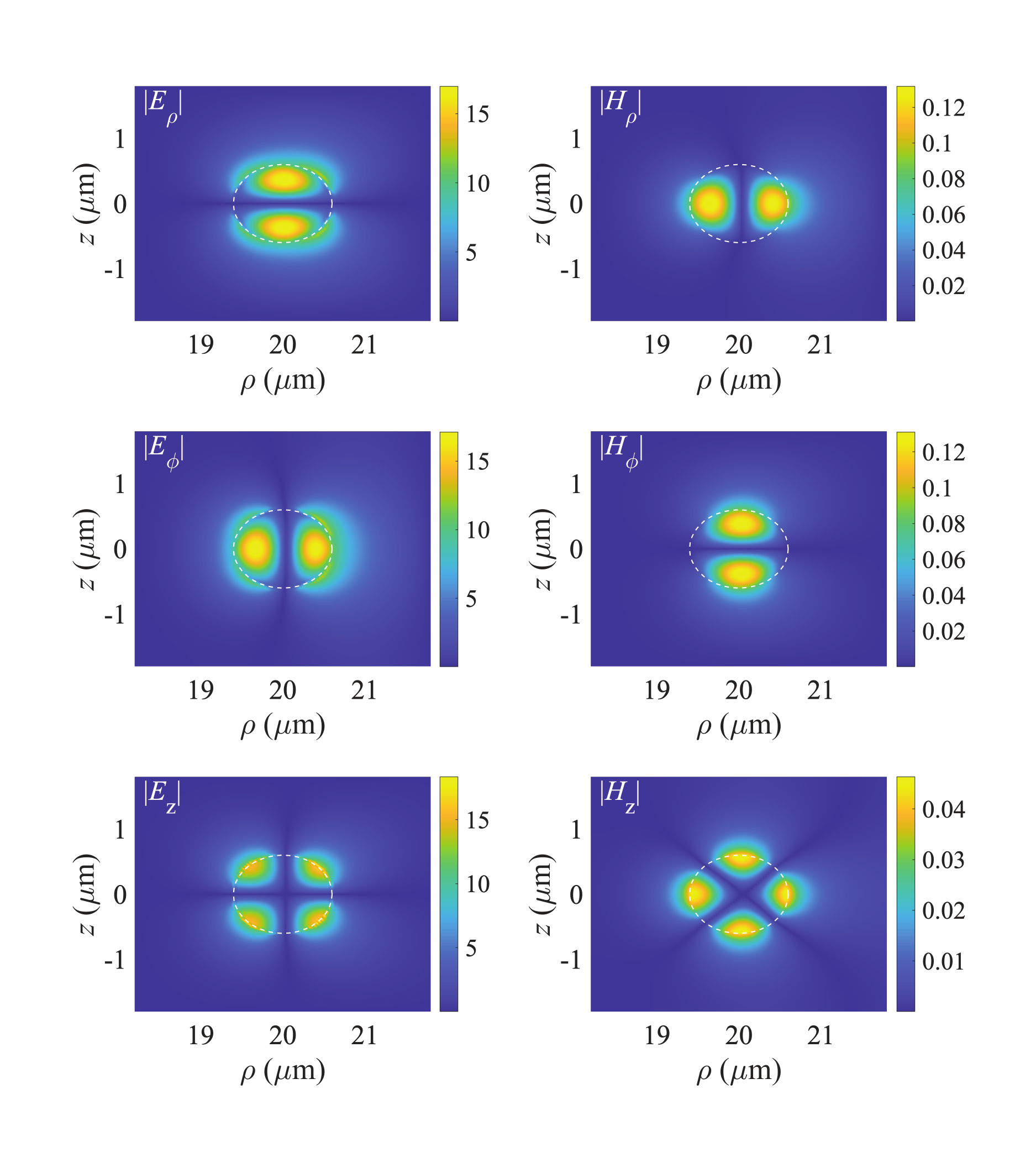}
    \caption{The magnitude of electric and magnetic field components for the fifth resonant modes inside a Si$_3$N$_4$ torus. The white dashed lines are added to visualize the location of the torus on the $\rho-z$ plane.}
    \label{fig:torus_fields}
\end{figure}

In Fig. \ref{fig:torus_fields}, we present the magnitude of the electric and magnetic field components for the fifth resonant mode within the torus. Notably, despite our solver being based on finite-difference computations over a rectangular grid, the resulting field profiles closely resemble second-order Bessel functions, which are characteristic eigensolutions of the resonant modes in circular waveguides.

\section{Conclusion}\label{section6}
We have developed and validated a robust numerical formulation for computing resonant modes in dielectric ring resonators embedded in cylindrically symmetric environments. The proposed method successfully addresses the challenge of modeling complex structures with inhomogeneous backgrounds by solving a two-dimensional eigenproblem derived from Maxwell's equations in cylindrical coordinates.

The key contributions of this work include:
\begin{itemize}
    \item The derivation of three complementary formulations (E-field, H-field, and mixed E-H) that provide flexibility in handling different boundary conditions and material interfaces
    \item An efficient numerical implementation using finite-difference discretization on a $\rho$-$z$ grid, transforming the problem into a manageable eigenvalue problem
    \item A two-step algorithm for accurately determining the azimuthal mode number $m$ and corresponding propagation constant $\beta$
    \item Comprehensive validation against established commercial solvers, demonstrating computational accuracy with differences typically below 0.3\% in effective refractive indices
\end{itemize}

The solver's capability was demonstrated through three representative examples: a buried Si$_3$N$_4$ ring showing strong field confinement, a ring on a substrate exhibiting field skewing effects, and a torus structure revealing complex modal patterns. In all cases, the computed field distributions and effective indices showed excellent agreement with reference solutions from COMSOL Multiphysics and other commercial tools.

\bibliographystyle{ieeetr}
\bibliography{references}

\end{document}